# Reduction of Second-Order Network Systems with Structure Preservation

Xiaodong Cheng, Yu Kawano, and Jacquelien M.A. Scherpen

**Abstract**

This paper proposes a general framework for structure-preserving model reduction of a second-order network system based on graph clustering. In this approach, vertex dynamics are captured by the transfer functions from inputs to individual states, and the dissimilarities of vertices are quantified by the $\mathcal{H}_2$-norms of the transfer function discrepancies. A greedy hierarchical clustering algorithm is proposed to place those vertices with similar dynamics into same clusters. Then, the reduced-order model is generated by the Petrov-Galerkin method, where the projection is formed by the characteristic matrix of the resulting network clustering. It is shown that the simplified system preserves an interconnection structure, i.e., it can be again interpreted as a second-order system evolving over a reduced graph. Furthermore, this paper generalizes the definition of network controllability Gramian to second-order network systems. Based on it, we develop an efficient method to compute $\mathcal{H}_2$-norms and derive the approximation error between the full-order and reduced-order models. Finally, the approach is illustrated by the example of a small-world network.

## I. INTRODUCTION

Network systems capture the behaviors and dynamics of many interconnected physical models and have received compelling attention from the system and control community, see [1]–[3] for an overview. A variety of network systems, such as distributed power grids or mass-damper-spring networks, are given as differential equations in second-order form, see [4], [5]. For

Xiaodong Cheng, Yu Kawano and Jacquelien M.A. Scherpen are with Jan C. Willems Center for Systems and Control, Engineering and Technology Institute Groningen, Faculty of Mathematics and Natural Sciences, University of Groningen, Nijenborgh 4, 9747 AG Groningen, the Netherlands. {x.cheng,y.kawano,j.m.a.scherpen}@rug.nl

large-scale networks, the second-order form dynamic models can be so complex and high-dimensional that system analysis and controller design become considerably difficult because of the impractical amounts of storage space and computation. Therefore, it is desirable to derive a lower-dimensional model which has an input-output behavior similar to the original one as well as inherits a second-order network structure. The preservation of the second-order realization allows an insightful physical interpretation. Moreover, the interconnection topology of the reduced model is important in the further applications, such as distributed controller designs and sensor allocations.

However, deriving reduced models for second-order network systems is not necessarily straightforward. Indeed, we are able to convert a second-order system to its equivalent first-order representation and then apply the reduction techniques used for first-order systems. However, the resulting models are not of second-order form in general. In [6]–[9] etc., the existing model reduction methods, including balanced truncation and moment matching, have been extended to the second-order case. Although the resulting reduced model is presented in a second-order form, it fails to preserve the interconnection topology among subsystems, i.e., such reduced models cannot be interpreted as network systems anymore.

There is another attempt to simplify the complexity of second-order networks based on time-scale separation and singular perturbation analysis, see e.g., [10] and references therein. In [10], this approach identifies the sparsely and densely connected areas of a power grids, and then aggregates the state variables of the coherent areas. Through singular perturbation approximation, the algebraic structure of Laplacian matrix is maintained, therefore, this approach indeed can preserve the network structure. However, this method does not explicitly consider the influence of the external inputs into the networks, and there is no analytical expression for the approximation error between the original and aggregated model.

Recently, clustering-based model reduction methods for first-order network systems have been investigated in [11]–[15]. The methods are interpreted as the Petrov-Galerkin approximation whose projection matrices are generated from some suitable clusterings (or partitions) of graphs. The advantage of the clustering-based approach is that the reduction process has a meaningful interpretation, and the interconnection structures are preserved in the reduced-order systems. However, seeking the finest graph clustering such that the reduced model achieves the smallest approximation error is an *NP-hard problem* [16], which may require almost impractical amounts of computation power for large-scale networks. Therefore, the most crucial step for model reduction is to exploit some efficient algorithms for cluster selection. To this end, different selection techniques are developed, see [12]–[15] and the references therein.

Paper [12] introduces the concept of an *almost equitable partition* (AEP). If the AEP of the underlying graph exists, then we are able to obtain an explicit $\mathcal{H}_2$ error expression between the original and simplified models. However, AEPs are difficult to find for general networks, therefore, the use of this approach is limited. The second method is proposed in [13] for reduction of network systems evolving over a tree graph. The so-called *generalized edge controllability and observability Gramains* are introduced to identify the importance of edges, and network clustering is then carried out by iteratively aggregating the vertices linked by the least important edge. However, the applicability of this method is also limited since the approximation process is restricted to the graphs with tree topology. Furthermore, the extension of this method to second-order networks is not straightforward. The approach in [14], [15] offers another feasible solution for network simplification, which is then extended to the second-order case in [17]. In their method, graph clustering is performed based on *cluster reducibility*, which is generalized as the uncontrollability of clusters and computed through a tridiagonal realization of their first-order representation. Then, the reducible clusters are merged to construct a reduced model with preservation of a second-order network topology. Nevertheless, this approach does not take the algebraic structure of the Laplacian matrix into account, and the approximation procedure and error analysis are reliant on the asymptotic stability of the system. This can be a limitation for some applications, e.g., the coupled swing dynamics in power networks as in [5], [10].

In this paper, we propose a novel model reduction approach for second-order network systems based on graph clustering. In contrast to the existing techniques, this method can be applied to more general network models, which do not restrict to a special class of partitions as in [12] or to a tree topology as in [13], or to asymptotically stable models as in [17]. Besides, unlike [10], we consider the system dynamics with respect to input signals. In [18], preliminary results are presented, which are generalized in this paper by extending the definition of controllability Gramian and proposing a new cluster algorithm. Specifically, the main contributions of this paper are as follows.

First, a clustering-based model reduction for second-order network system is proposed in the framework of Galerkin projection. The characteristic matrix of a graph clustering is used as the projection so that the interconnection topology can be preserved in the reduced-order model. More importantly, the algebraic structure of Laplacian matrix is also retained, and consequently, the reduced graph can be reconstructed.

Second, we design a greedy hierarchical clustering algorithm to generate an appropriate network partition. Specifically, we characterize the behaviors of vertices by the transfer functions from inputs to their individual states and denote the dissimilarities by the $\mathcal{H}_2$-norms of the transfer function deviations. Then, a systematic process places those vertices with almost similar behaviors into same clusters. The feasibility and efficiency of this method are demonstrated by a numerical example.

Third, the *network controllability Gramian* is generalized, which provides an efficient approach to quantify the dissimilarities of vertices in terms of $\mathcal{H}_2$-norms. Since the network systems are not necessarily asymptotically stable, the conventional definitions of Gramians in e.g., [19] are not applicable here. In [20], we propose the novel definition of network Gramian, which is associated with a Lyapunov-like equation. Here, we extend this definition to the second-order case using the synchronization properties. Furthermore, the approximation error between the full-order and reduced-order systems is also derived from the network controllability Gramian of an error system.

The remainder of this paper is organized as follows. Section II presents the mathematical model of second-order network systems and formulates the problem of structure-preserving model reduction. In Section III, we provide the framework of clustering-based model reduction. Then, in Section IV, we discuss the network controllability Gramian and propose the cluster selection algorithm. Finally, Section V illustrates the feasibility of our method by means of an numerical example, and Section VI concludes the whole paper.

*Notation:* $A$ is *semistable* if and only if $A$ is Lyapunov stable and $A$ has no nonzero imaginary eigenvalues [21]. $A \sim B$ means that the square matrices $A$ and $B$ are *similar*, i.e., they are related by a similarity transformation that is $B = T^{-1}AT$ with $T$ a square nonsingular matrix. A *block diagonal matrix*, denoted by $\text{diag}(A_1, \cdots, A_n)$, is a square diagonal matrix in which the diagonal elements are square matrices $A_i$, and the off-diagonal elements are 0. Besides, we use the following notation throughout this paper, where the subscript $n$ of $I_n$ or $\mathbf{1}_n$ is omitted when no confusion arises.

## II. PROBLEM FORMULATION

An undirected connected graph is defined by a pair $\mathcal{G} = (\mathcal{V}, \mathcal{E})$, where $\mathcal{V}$ and $\mathcal{E} \subseteq \mathcal{V} \times \mathcal{V}$ represent the sets of vertices and edges, respectively. Assume that $|\mathcal{V}| = n$ and $|\mathcal{E}| = n_\text{e}$, then

| | |
|---|---|
| $\mathbb{R}$ | set of real numbers |
| $I_n$ | identity matrix of size $n$ |
| $\mathbf{e}_i$ | $i$-th column vector of $I_n$ |
| $\mathbf{e}_{ij}$ | $\mathbf{e}_i - \mathbf{e}_j$ |
| $\mathbf{1}_n$ | all-ones vector of $n$ entries |
| $\|A\|_2$ | induced 2-norm of matrix $A$ |
| $\mathbf{rank}(A)$ | rank of matrix $A$ |
| $\mathbf{tr}(A)$ | trace of matrix $A$ |
| $\mathrm{span}(v_1, \cdots, v_n)$ | span of a set of vectors $\{v_1, \cdots, v_n\}$ |
| $|\mathcal{S}|$ | cardinality of set $\mathcal{S}$ |
| $\|T(s)\|_{\mathcal{H}_2}$ | $\mathcal{H}_2$-norm of transfer function $T(s)$ |
| $\|T(s)\|_{\mathcal{H}_\infty}$ | $\mathcal{H}_\infty$-norm of transfer function $T(s)$ |
| $A \succ 0 \; (A \prec 0)$ | positive (negative) definiteness of a symmetric matrix $A$ |
| $A \succcurlyeq 0 \; (A \preccurlyeq 0)$ | positive (negative) semi-definiteness of a symmetric matrix $A$ |

the *incidence matrix* of $\mathcal{G}$ is defined by $R \in \mathbb{R}^{n \times n_e}$ such that $R_{ij} = 1$ if the edge $(i,j)$ heads to vertex $i$, $-1$ if it leaves vertex $i$ and $0$ otherwise. For an undirected graph, $R$ can be obtained by assigning each edge with an arbitrary orientation. The *weighted Laplacian matrix* of graph $\mathcal{G}$, denoted by $L \in \mathbb{R}^{n \times n}$, is defined by

$$L = RWR^T, \tag{1}$$

where $W \in \mathbb{R}^{n_e \times n_e}$ is the diagonal positive definite matrix whose diagonal entries represent the weights of edges.

Consider a network system evolving over graph $\mathcal{G}$, which has a linear time-invariant description in second-order form as

$$\boldsymbol{\Sigma}: M\ddot{x} + D\dot{x} + Lx = Fu, \tag{2}$$

where $x \in \mathbb{R}^n$ and $u \in \mathbb{R}^m$ denote the vertex states and external inputs, respectively. In this model, $M, D, L \in \mathbb{R}^{n \times n}$ are referred to inertia, damping and stiffness matrices, respectively.

A variety of physical network systems are modeled in the form of (2), including the linearized swing equation in power grids [5] and mass-damper-spring networks [4]. Take the latter one for instance, $M$ represents the distribution of masses, and $D$ presents the dampers on edges and vertices, while $L$ indicates the strength of diffusive coupling among the vertices connected by springs. Based on practical applications, the following *structural conditions* are assumed.

**Assumption 1.** ① $M \succ 0$ *is diagonal;* ② $D = D^T \succ 0$; ③ $L = L^T \succcurlyeq 0$ *is a weighted Laplacian matrix of a connected undirected graph. (For the properties of L, we refer to [22]).*

Assumption 1 guarantees that the system $\Sigma$ is *passive* and *semistable*, which is shown by the following reasoning.

First, the total energy of $\Sigma$ is given by

$$H(x, \dot{x}) = \frac{1}{2}\dot{x}^T M \dot{x} + \frac{1}{2} x^T L x. \tag{3}$$

With $y = F^T \dot{x}$ as output, we have

$$u^T y - \dot{H} = u^T F^T \dot{x} - \dot{x}^T M \ddot{x} - x^T L \dot{x}$$
$$= u^T F^T \dot{x} - \dot{x}^T(-D\dot{x} - Lx + Fu) - x^T L \dot{x} = \dot{x}^T D \dot{x} > 0.$$

It follows from [23] that the system $\Sigma$ is passive. Moreover, $\Sigma$ can be presented in the form of a port-Hamiltonian system as in [24].

Second, the stability of the system $\Sigma$ can be seen from the first-order form realization

$$\dot{\mathcal{X}} = \mathcal{A}\mathcal{X} + \mathcal{B}u \tag{4}$$

with $\mathcal{X}^T = \begin{bmatrix} x^T, \dot{x}^T \end{bmatrix}$ as the $2n$-dimensional state and

$$\mathcal{A} = \begin{bmatrix} \mathbf{0}_{n \times n} & I \\ -M^{-1}L & -M^{-1}D \end{bmatrix}, \ \mathcal{B} = \begin{bmatrix} \mathbf{0}_{n \times m} \\ M^{-1}F \end{bmatrix}. \tag{5}$$

From Assumption 1, it is easy to check that all the eigenvalues of $\mathcal{A}$ are in the closed left-half plane, and only one of them is at the origin. Therefore, the second-order network system $\Sigma$ is semistable.

**Remark 1.** *The major differences with set-ups from familiar second-order systems as in [6]–[8] are that $D$ and $L$ in (2) contain the information of network spatial structures, and the system $\Sigma$ is not asymptotically stable.*

Now we formulate the problem of model reduction for second-order network systems as follows.

**Problem 1.** *Given a second-order network system $\Sigma$ as in (2), find a pair of projection matrices $W, V \in \mathbb{R}^{n \times r}$ with $r \ll n$ to construct a reduced model in the second-order form as*

$$\hat{\Sigma} : \begin{cases} \hat{M}\ddot{z} + \hat{D}\dot{z} + \hat{L}z = W^T F u, \\ \hat{x} = Vz, \end{cases} \quad (6)$$

*where $\hat{M} = W^T M V$, $\hat{D} = W^T D V$ and $\hat{L} = W^T L V \in \mathbb{R}^{r \times r}$. We require the matrices $\hat{M}$, $\hat{D}$ and $\hat{L}$ to fulfill the **structural conditions** in Assumption 1 and the trajectories of $\hat{x}(t)$ to approximate those of $x(t)$ in the original system $\Sigma$ with a small error.*

We call Problem 1 a *position-based model reduction* for second-order network systems, since the variable $\hat{x}(t)$ in (6) is used to approximate $x(t)$ rather than $\dot{x}(t)$. However, Problem 1 can be easily modified to solve *velocity-based model reduction* problems, where the second equation in (6) is replaced by $v = V\dot{z}$, and it requires $v$ and $\dot{x}(t)$ to have close behaviors respect to the external input fluxes $u(t)$.

In this paper, we first consider the position-based model reduction as a standard problem setting for network systems and discuss the solution of Problem 1 in the following two sections. Besides, we briefly state the extension of our proposed method to velocity-based model reduction.

## III. CLUSTERING-BASED MODEL REDUCTION

This section will first give a class of Galerkin projections that can deliver reduced second-order network models with interconnection structures. Then, some important properties of the resulting systems is discussed.

Before proceeding, we recapitulate the notions of network clustering and its characteristic matrix from e.g., [25].

**Definition 1.** *Consider a connected graph $\mathcal{G} = (\mathcal{V}, \mathcal{E})$, where $\mathcal{V} = \{1, 2, \cdots, n\}$ is the index set of vertices. A nonempty index subset of $\mathcal{V}$, denoted by $\mathcal{C}$, is called a **cluster** of graph $\mathcal{G}$. Then, **network clustering** is to partition $\mathcal{V}$ into $r$ disjoint clusters which cover all the elements in $\mathcal{V}$.*

**Definition 2.** *Consider a network clustering $\{\mathcal{C}_1, \mathcal{C}_2, \cdots, \mathcal{C}_r\}$ of a vertex set $\mathcal{V}$ with $|\mathcal{V}| = n$. The **characteristic vector** of the cluster $\mathcal{C}_i$ is defined by binary vector $p(\mathcal{C}_i) \in \mathbb{R}^n$ where $\mathbf{1}_n^T p(\mathcal{C}_i) =$*

$|\mathcal{C}_i|$, and the $k$-th element of $p(\mathcal{C}_i)$ is 1 when $k \in \mathcal{C}_i$ and 0 otherwise. Then, **characteristic matrix** of the clustering is a binary matrix defined by

$$P := [p(\mathcal{C}_1), p(\mathcal{C}_2), \cdots, p(\mathcal{C}_r)] \in \mathbb{R}^{n \times r}. \tag{7}$$

Now, consider a network system $\Sigma$ on graph $\mathcal{G}$ with $n$ vertices. To approximate $\Sigma$ by an $r$-th dimensional reduced model, we need to find a network clustering which partitions $n$ vertices into $r$ clusters. To preserve the structural conditions, we then characterize the projection in Problem 1 by the characteristic matrix of a graph clustering. More precisely, the following unnormalized Galerkin projection is applied

$$W = V = P, \tag{8}$$

which leads to the $r$-dimensional reduced second-order network system as

$$\hat{\Sigma} : \begin{cases} \hat{M}\ddot{z} + \hat{D}\dot{z} + \hat{L}z = P^T F u, \\ \hat{x} = Pz, \end{cases} \tag{9}$$

where $\hat{M} = P^T M P$, $\hat{D} = P^T D P$ and $\hat{L} = P^T L P \in \mathbb{R}^{r \times r}$.

This projection also can be found in [11], [22], [26], [27]. In this paper, we will further discuss this idea and develop our model reduction method based on it. The following proposition holds for the simplified model in (9).

**Proposition 1.** *The reduced network system $\hat{\Sigma}$ resulting from a clustering-based projection as in (8) preserves the interconnection structures of the original system $\Sigma$, i.e., $\hat{M}$, $\hat{D}$ and $\hat{L}$ satisfy the **structural conditions** in Assumption 1.*

*Proof.* Observe that $P$ is a binary matrix with full column rank. It is not hard to verify that $\hat{M} \succ 0$, $\hat{D} \succ 0$ and $\hat{L} \succcurlyeq 0$.

Furthermore, since there always exists a permutation matrix $T$ such that

$$\tilde{P} := TP = \text{diag}(\mathbf{1}_{|\mathcal{C}_1|}, \mathbf{1}_{|\mathcal{C}_2|}, \cdots, \mathbf{1}_{|\mathcal{C}_r|}), \tag{10}$$

we have $P = T^T \tilde{P}$ and $\hat{M} = \tilde{P}^T T M T^T \tilde{P}$. Clearly, matrix $T M T^T$ is diagonal, and therefore, $\hat{M}$ is also. Moreover, the $i$-th diagonal entry of $\hat{M}$ presents the sum of all the masses with the $i$-th cluster.

From definition (1), we have $\hat{L} = P^T R W R^T P$. Suppose the edge $(i,j)$ of the original graph is represented by $R_k$, the $k$-th column of the incidence matrix $R$. Then, the entries of $R_k$ satisfy that $R_{i,k} = -R_{j,k}$, and the other entries are zero. If vertices $i$ and $j$ are within the same cluster, from the definition of characteristic matrix of clustering, we have $P_i = P_j$, where $P_i$ is the $i$-th row of $P$. Hence, we obtain $P^T R_k = 0$. Furthermore, we can define a new incidence matrix $\hat{R}$ by removing all the zero columns of $P^T R$ and a new edge weight matrix $\hat{W}$ by eliminating the rows and columns which are corresponding to the edges linking vertices in a same cluster. Consequently, it follows that $\hat{L} = P^T R W R^T P = \hat{R} \hat{W} \hat{R}^T$, where $\hat{L}$ is also a Laplacian matrix of an undirected connected graph. □

From the algebraic structures of matrices $\hat{M}$, $\hat{D}$ and $\hat{L}$, we are able to reconstruct the topology of the reduced network. The following example then illustrates the intuitive interpretation of clustering-based model reduction.

**Example 1.** *The left inset of Fig. 1 depicts a mass-damper-spring network system of 4 vertices. The coefficient matrices are given by*

$$M = \begin{bmatrix} 1 & 0 & 0 & 0 \\ 0 & 2 & 0 & 0 \\ 0 & 0 & 1 & 0 \\ 0 & 0 & 0 & 2 \end{bmatrix}, D = \begin{bmatrix} 4 & -2 & 0 & -1 \\ -2 & 2 & 0 & 0 \\ 0 & 0 & 3.5 & -3 \\ -1 & 0 & -3 & 4 \end{bmatrix}, L = \begin{bmatrix} 4 & -1 & -2 & -1 \\ -1 & 3 & -1 & -1 \\ -2 & -1 & 5 & -2 \\ -1 & -1 & -2 & 4 \end{bmatrix}, F = \begin{bmatrix} 1 & 0 \\ 0 & 0 \\ 0 & 0 \\ 0 & 1 \end{bmatrix}.$$

*If vertex 3 and 4 are clustered, i.e., the network clustering is $\{\{1\}, \{2\}, \{3,4\}\}$, the characteristic matrix $P$ is then generated as*

$$P = \begin{bmatrix} 1 & 0 & 0 \\ 0 & 1 & 0 \\ 0 & 0 & 1 \\ 0 & 0 & 1 \end{bmatrix}, \tag{11}$$

*which leads to the reduced-order model as*

$$\hat{M} = \begin{bmatrix} 1 & 0 & 0 \\ 0 & 2 & 0 \\ 0 & 0 & 3 \end{bmatrix}, \hat{D} = \begin{bmatrix} 4 & -2 & -1 \\ -2 & 2 & 0 \\ -1 & 0 & 1.5 \end{bmatrix}, \hat{L} = \begin{bmatrix} 4 & -1 & -3 \\ -1 & 3 & -2 \\ -3 & -2 & 5 \end{bmatrix}, P^T F = \begin{bmatrix} 1 & 0 \\ 0 & 0 \\ 0 & 1 \end{bmatrix}.$$

*Clearly, the algebraic structures of the inertia, damper, and stiffness matrices are preserved in the new system. An interpretation of the reduced model is presented in the right inset of Fig. 1.*

Next, we discuss some important properties of the reduced second-order network system (9). First, the following proposition can be easily obtained.

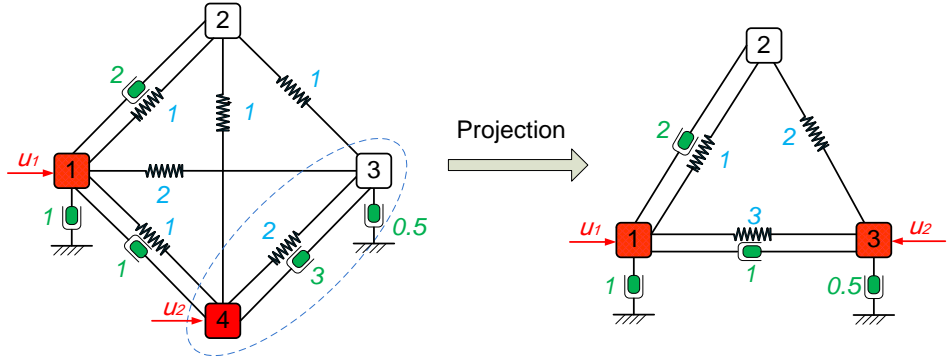

Fig. 1: An illustrative example of clustering-based model reduction for a mass-damper-spring network system, where red blocks represent the controlled vertices.

**Proposition 2.** *The reduced second-order network system $\hat{\Sigma}$ in (9) preserves the semistability and passivity of the original system $\Sigma$.*

*Proof.* Proposition 1 states that the reduced matrices $\hat{M}$, $\hat{D}$ and $\hat{L}$ also fulfill the structural conditions list in Assumption 1. Therefore, we can also show the passivity and semistability of the reduced model $\hat{\Sigma}$ by converting it to a first-order realization. □

Second, the synchronization properties of the original network system $\Sigma$ is also retained in the reduced model $\hat{\Sigma}$. The following theorem is extended from [22] and [26], where first-order network systems are studied.

**Theorem 1.** *Consider the second-order network system $\Sigma$ and its reduced model $\hat{\Sigma}$ resulting from an clustering-based projection. If their initial conditions satisfy $x(0) = Pz(0) = \hat{x}(0)$ and $\dot{x}(0) = P\dot{z}(0) = \dot{\hat{x}}(0)$, then the trajectories of both systems with $u = 0$ converge to a common value. More precisely,*

$$\lim_{t\to\infty} x(t) = \lim_{t\to\infty} \hat{x}(t) = \sigma_D^{-1} \begin{bmatrix} \mathbf{1}\mathbf{1}^T \left(Dx(0) + M\dot{x}(0)\right) \\ \mathbf{0}_{n\times 1} \end{bmatrix},$$

$$\lim_{t\to\infty} \dot{x}(t) = \lim_{t\to\infty} \dot{\hat{x}}(t) = 0,$$

*where $\sigma_D = \mathbf{1}^T D \mathbf{1}$.*

*Proof.* First, the synchronization of $\Sigma$ is proved as follows.

Since $\mathcal{A}$ in (4) has one zero eigenvalue, we consider the Jordan matrix decomposition $\mathcal{A} = \mathsf{U}\Lambda\mathsf{U}^{-1}$, where

$$\Lambda = \begin{bmatrix} 0 & \\ & \bar{\Lambda} \end{bmatrix}, \tag{12}$$

with $\bar{\Lambda}$ Hurwitz. The first row of $\mathsf{U}^{-1}$ and the first column of $\mathsf{U}$, denoted by $\mathsf{v}_1 \in \mathbb{R}^{1 \times 2n}$ and $\mathsf{u}_1 \in \mathbb{R}^{2n \times 1}$, are the left and right eigenvectors of $\mathcal{A}$ corresponding to the only zero eigenvalue, respectively. Here, $\mathsf{u}_1$ is an unit vector, and we have

$$\mathcal{A}^T \mathsf{v}_1^T = 0, \quad \mathcal{A}\mathsf{u}_1 = 0 \text{ and } \mathsf{v}_1 \mathsf{u}_1 = 1. \tag{13}$$

The above equations then lead to

$$\mathsf{v}_1 = \frac{\sqrt{n}}{\mathbf{1}^T D \mathbf{1}} \begin{bmatrix} \mathbf{1}^T D & \mathbf{1}^T M \end{bmatrix}, \text{ and } \mathsf{u}_1 = \begin{bmatrix} \frac{1}{\sqrt{n}}\mathbf{1} \\ \mathbf{0}_{n \times 1} \end{bmatrix}, \tag{14}$$

where the property $L\mathbf{1} = 0$ is used. Furthermore, we partition $\mathsf{U}$ and $\mathsf{U}^{-1}$ as

$$\mathsf{U} = \begin{bmatrix} \mathsf{u}_1 & \mathsf{U}_2 \end{bmatrix}, \quad \mathsf{U}^{-1} = \begin{bmatrix} \mathsf{v}_1 \\ \mathsf{V}_2 \end{bmatrix}, \tag{15}$$

and it follows that

$$e^{\mathcal{A}t} = \mathsf{U}e^{\Lambda t}\mathsf{U}^{-1} = \begin{bmatrix} \mathsf{u}_1 & \mathsf{U}_2 \end{bmatrix} \begin{bmatrix} 1 & \\ & e^{\bar{\Lambda}t} \end{bmatrix} \begin{bmatrix} \mathsf{v}_1 \\ \mathsf{V}_2 \end{bmatrix} = \mathsf{u}_1\mathsf{v}_1 + \mathsf{U}_2 e^{\bar{\Lambda}t} \mathsf{V}_2. \tag{16}$$

Observe that $\lim_{t \to \infty} \mathsf{U}_2 e^{\bar{\Lambda}t} \mathsf{V}_2 = 0$, which yields that

$$\lim_{t \to \infty} e^{\mathcal{A}t} = \mathsf{u}_1 \mathsf{v}_1 = \sigma_D^{-1} \begin{bmatrix} \mathbf{1}_n \mathbf{1}_n^T D & \mathbf{1}_n \mathbf{1}_n^T M \\ \mathbf{0}_{n \times n} & \mathbf{0}_{n \times n} \end{bmatrix}. \tag{17}$$

Consequently, we obtain

$$\lim_{t \to \infty} \begin{bmatrix} x(t) \\ \dot{x}(t) \end{bmatrix} = \lim_{t \to \infty} e^{\mathcal{A}t} \begin{bmatrix} x(0) \\ \dot{x}(0) \end{bmatrix} = \sigma_D^{-1} \begin{bmatrix} \mathbf{1}_n \mathbf{1}_n^T \left( Dx(0) + M\dot{x}(0) \right) \\ \mathbf{0}_{n \times 1} \end{bmatrix}.$$

Proposition 1 indicates that the reduced-order model $\hat{\Sigma}$ has the same form as the original model $\Sigma$. Therefore, a similar reasoning line yields

$$\lim_{t \to \infty} \begin{bmatrix} z(t) \\ \dot{z}(t) \end{bmatrix} = \frac{1}{\mathbf{1}_r^T \hat{D} \mathbf{1}_r} \begin{bmatrix} \mathbf{1}_r \mathbf{1}_r^T \left( \hat{D}z(0) + \hat{M}\dot{z}(0) \right) \\ \mathbf{0}_{r \times 1} \end{bmatrix},$$

Then, we have the following equations for $\hat{\Sigma}$:

$$\lim_{t\to\infty} \dot{\hat{x}}(t) = \lim_{t\to\infty} P\dot{z}(t) = \mathbf{0}_{n\times 1},$$

$$\lim_{t\to\infty} \hat{x}(t) = \lim_{t\to\infty} Pz(t) = P\frac{\mathbf{1}_r\mathbf{1}_r^T\left(\hat{D}z(0) + \hat{M}\dot{z}(0)\right)}{\mathbf{1}_r^T\hat{D}\mathbf{1}_r},$$

where $\hat{M} = P^T MP$, $\hat{D} = P^T DP$. The result immediately follows from $\hat{x}(0) = Pz(0) = x(0)$ and $P\mathbf{1}_r = \mathbf{1}_n$. $\square$

**Remark 2.** *Theorem 1 also indicates that with any initial conditions and $u = 0$, the trajectories of $\dot{x}(t)$ and $\dot{z}(t)$ always converge to zero, while those of $x(t)$ and $z(t)$ converge to a common value which is nonzero in general. Furthermore, if we denote $\xi(t)$ and $\hat{\xi}(t) \in \mathbb{R}^{n\times p}$ as the **impulse responses** of $\Sigma$ and $\hat{\Sigma}$, then we can obtain*

$$\begin{aligned}\lim_{t\to\infty} \xi(t) &= \lim_{t\to\infty} \hat{\xi}(t) = \sigma_D^{-1}\mathbf{1}\mathbf{1}^T F, \\ \lim_{t\to\infty} \dot{\xi}(t) &= \lim_{t\to\infty} \dot{\hat{\xi}}(t) = 0,\end{aligned} \qquad (18)$$

*which follows from the computation of $\lim_{t\to\infty} e^{\mathcal{A}t}\mathcal{B}$ with $\mathcal{A}$ and $\mathcal{B}$ defined in (5).*

## IV. SELECTION OF NETWORK CLUSTERING

The cluster selection is a crucial problem in clustering-based model reduction, since different choices of clustering yield different reduced models with different approximation qualities. In Section IV-A, we first introduce the concept of the *second-order network controllability Gramian* and then discuss a method to compute such Gramian from the state-space model (4). The purpose of defining such a Gramian is explained in Section IV-B, where a hierarchical clustering algorithm is designed for cluster selection. Finally, the approximation error is analyzed in Section IV-C.

### A. Second-order Network Controllability Gramian

The conventional definition of controllability Gramian (see [19], [28]) is given by

$$\mathcal{P}_s = \int_0^\infty e^{\mathcal{A}t}\mathcal{B}\mathcal{B}^T e^{\mathcal{A}^T t}dt \qquad (19)$$

with $\mathcal{A}$ and $\mathcal{B}$ the coefficient matrices in the state-space realization (4). However, this standard definition is not applicable for the network system $\Sigma$, since it is restricted to asymptotically stable systems. The integral in (19) is not well-defined for semistable systems, because the impulse response $e^{\mathcal{A}t}\mathcal{B}$ does not necessarily converge to zero as $t \to \infty$. In [20], we have proposed the definition of the network controllability Gramian for semistable first-order systems. In this paper, we extend this notation to the second-order case. First, the definition of *convergence matrix* is introduced as follows.

**Definition 3.** *The **convergence matrix** of the state-space representation in (4) is defined by*

$$\mathcal{J} = \lim_{t \to \infty} e^{\mathcal{A}t} \in \mathbb{R}^{2n \times 2n}. \tag{20}$$

For the second-order network system $\Sigma$, we obtain a concise expression of the convergence matrix from (17) as

$$\mathcal{J} = \sigma_D^{-1} \begin{bmatrix} \mathbf{1}\mathbf{1}^T D & \mathbf{1}\mathbf{1}^T M \\ \mathbf{0}_{n \times n} & \mathbf{0}_{n \times n} \end{bmatrix}. \tag{21}$$

Based on this, we have the following definition of a new Gramian matrix.

**Definition 4.** *Consider the second-order network system $\Sigma$ in (2) and the coefficient matrix pair $(\mathcal{A}, \mathcal{B})$ defined in (4). Then, the **second-order network controllability Gramian** of the system $\Sigma$, denoted by $\mathcal{P} \in \mathbb{R}^{2n \times 2n}$, is defined by*

$$\mathcal{P} = \int_0^\infty (e^{\mathcal{A}t} - \mathcal{J})\mathcal{B}\mathcal{B}^T(e^{\mathcal{A}^T t} - \mathcal{J}^T) dt. \tag{22}$$

Notice that, in general, $\mathcal{P}$ is positive semidefinite, which will be explained later. Next, we explore a method to find the $\mathcal{P}$ without computing the integral in (22). The following theorem links the network Gramian with the solutions of a Lyapunov-like equation.

**Theorem 2.** *The second-order network controllability Gramian $\mathcal{P}$ of the system $\Sigma$ is a solution of the following linear matrix equation*

$$\mathcal{A}\tilde{\mathcal{P}} + \tilde{\mathcal{P}}\mathcal{A}^T + (I - \mathcal{J})\mathcal{B}\mathcal{B}^T(I - \mathcal{J}^T) = 0, \tag{23}$$

*where $\mathcal{J}$ is the convergence matrix of $\Sigma$.*

*Proof.* By the definition of $\mathcal{J}$, we first have the integral

$$\int_0^\infty \frac{d}{dt} \left[ (e^{\mathcal{A}t} - \mathcal{J})\mathcal{B}\mathcal{B}^T(e^{\mathcal{A}^T t} - \mathcal{J}^T) \right] dt \tag{24}$$
$$= \left[ (e^{\mathcal{A}t} - \mathcal{J})\mathcal{B}\mathcal{B}^T(e^{\mathcal{A}^T t} - \mathcal{J}^T) \right] \Big|_0^\infty = -(I - \mathcal{J})\mathcal{B}\mathcal{B}^T(I - \mathcal{J}^T).$$

Furthermore, we have

$$\frac{d}{dt}\left[(e^{\mathcal{A}t} - \mathcal{J})\mathcal{B}\mathcal{B}^T(e^{\mathcal{A}^T t} - \mathcal{J}^T)\right] \tag{25}$$
$$= \mathcal{A}e^{\mathcal{A}t}\mathcal{B}\mathcal{B}^T(e^{\mathcal{A}^T t} - \mathcal{J}^T) + (e^{\mathcal{A}t} - \mathcal{J})\mathcal{B}\mathcal{B}^T e^{\mathcal{A}^T t}A^T.$$

The property $L\mathbf{1} = 0$ yields

$$\mathcal{AJ} = \sigma_D^{-1} \begin{bmatrix} \mathbf{0} & I \\ -M^{-1}L & -M^{-1}D \end{bmatrix} \begin{bmatrix} \mathbf{1}\mathbf{1}^T D & \mathbf{1}\mathbf{1}^T M \\ \mathbf{0}_{n\times n} & \mathbf{0}_{n\times n} \end{bmatrix} = 0.$$

Then, we obtain

$$\begin{aligned}\int_0^\infty \mathcal{A}e^{\mathcal{A}t}\mathcal{B}\mathcal{B}^T(e^{\mathcal{A}^T t} - \mathcal{J}^T)dt &= \mathcal{A}\int_0^\infty (e^{\mathcal{A}t} - \mathcal{J} + \mathcal{J})\mathcal{B}\mathcal{B}^T(e^{\mathcal{A}^T t} - \mathcal{J}^T)dt \\ &= \mathcal{A}\int_0^\infty (e^{\mathcal{A}t} - \mathcal{J})\mathcal{B}\mathcal{B}^T(e^{\mathcal{A}^T t} - \mathcal{J}^T)dt = \mathcal{A}\mathcal{P},\end{aligned} \quad (26)$$

and similarly,

$$\int_0^\infty (e^{\mathcal{A}t} - \mathcal{J})\mathcal{B}\mathcal{B}^T e^{\mathcal{A}^T t}\mathcal{A}^T dt = \mathcal{P}\mathcal{A}^T. \quad (27)$$

Finally, we obtain (23) by combining (24), (26) and (27). □

**Remark 3.** *It is worth emphasizing that the definition of the Gramian in (22) is a generalization of the conventional definition in (19). If the Laplacian matrix $L$ in (2) is replaced by a positive definite matrix (e.g. the network model considered in [17]), then $\mathcal{A}$ in (4) is Hurwitz, and the convergence matrix $\mathcal{J}$ in (20) is zero. In this case, it is clear that the network controllability Gramian $\mathcal{P}$ is equivalent to the standard definition of controllability Gramian $\mathcal{P}_s$ in (19). Moreover, equation (23) is reduced to the standard Lyapunov equation*

$$\mathcal{A}\mathcal{P} + \mathcal{P}\mathcal{A}^T + \mathcal{B}\mathcal{B}^T = 0, \quad (28)$$

*which has an unique solution.*

However, due to the semistability of $\Sigma$, the solution of the Lyapunov-like equation in (22) is not unique. This conclusion can be seen from the following lemma which provides sufficient and necessary conditions for the solution uniqueness of Sylvester equations.

**Lemma 1.** *[29] The continuous-time Sylvester equation is given by*

$$AX + XB = C, \quad (29)$$

*where $A$ and $B$ are real square matrices of sizes $n$ and $m$ respectively, and $X, C \in \mathbb{R}^{n\times m}$. Then, equation (29) has an unique solution $X$ for all $C$ if and only if $A$ and $-B$ share no eigenvalues.*

Observe that equation (23) is also a Sylvester equation and because $\mathcal{A}$ and $-\mathcal{A}^T$ share one zero eigenvalue, the solutions of equation (23) are not unique. To determine the second-order network controllability Gramian $\mathcal{P}$ from all the solutions of equation (23), we first find a relation between those solutions in the following lemma.

**Lemma 2.** *Suppose a symmetric matrix $\mathcal{P}_1 \in \mathbb{R}^{2n \times 2n}$ is a solution of the Lyapunov-like equation in (23), then the following three conditions are equivalent*

1) *A symmetric matrix $\mathcal{P}_2 \in \mathbb{R}^{2n \times 2n}$ is a solution of (23).*
2) *$\mathcal{P}_2$ satisfies the equation*
$$\mathcal{J}(\mathcal{P}_1 - \mathcal{P}_2)\mathcal{J}^T = \mathcal{P}_1 - \mathcal{P}_2. \tag{30}$$
3) *$\mathcal{P}_2$ can be expressed by*
$$\mathcal{P}_2 = \mathcal{P}_1 + \beta \Pi \tag{31}$$
*with $\Pi := \begin{bmatrix} \mathbf{1}\mathbf{1}^T & \mathbf{0}_{n \times n} \\ \mathbf{0}_{n \times n} & \mathbf{0}_{n \times n} \end{bmatrix}$ and $\beta$ a scalar constant.*

*Proof.* 1) $\Rightarrow$ 2): Assume that $\mathcal{P}_2$ is also a solution of (23), and denote $\Delta := \mathcal{P}_1 - \mathcal{P}_2$. We have
$$\mathcal{A}\Delta + \Delta \mathcal{A}^T = 0, \tag{32}$$
which leads to
$$e^{\mathcal{A}t}\left[\mathcal{A}\Delta + \Delta \mathcal{A}^T\right]e^{\mathcal{A}^T t} = \frac{d}{dt}\left[e^{\mathcal{A}t}\Delta e^{\mathcal{A}^T t}\right] = 0. \tag{33}$$
Therefore, we obtain
$$\int_0^\infty \frac{d}{dt}\left[e^{\mathcal{A}t}\Delta e^{\mathcal{A}^T t}\right] dt = 0, \tag{34}$$
which is equivalent to
$$\mathcal{J}\Delta \mathcal{J}^T = \Delta. \tag{35}$$

2) $\Rightarrow$ 3): Assume that $\mathcal{P}_2$ satisfies (30), i.e., equation (35) holds. Then, the entry in the $i$-th row and $j$-th column of $\Delta$ is given by $\Delta_{ij} = \mathcal{J}_i \Delta \mathcal{J}_j^T$, where $\mathcal{J}_i \in \mathbb{R}^{1 \times 2n}$ presents the $i$-th row of $\mathcal{J}$. Note that
$$\mathcal{J}_i = \begin{cases} \sigma_D^{-1} \cdot \left[\mathbf{1}^T D, \mathbf{1}^T M\right], & 1 \leq i \leq n \\ \mathbf{0}_{1 \times 2n}, & n+1 \leq i \leq 2n \end{cases} \tag{36}$$
with scalar $\sigma_D = \mathbf{1}^T D \mathbf{1}$. Therefore,
$$\Delta_{ij} = \sigma_D^{-2} \cdot \left(\mathbf{1}^T D \Delta D \mathbf{1} + \mathbf{1}^T M \Delta M \mathbf{1}\right) := \beta, \tag{37}$$

if $1 \leq i, j \leq n$, $\Delta_{ij} = 0$, otherwise, which implies that $\mathcal{P}_2$ is presented as equation (31).

3) $\Rightarrow$ 1): Observe that

$$\mathcal{A}\Pi = \begin{bmatrix} \mathbf{0} & I \\ -M^{-1}L & -M^{-1}D \end{bmatrix} \begin{bmatrix} \beta \mathbf{1}\mathbf{1}^T & \mathbf{0}_{n \times n} \\ \mathbf{0}_{n \times n} & \mathbf{0}_{n \times n} \end{bmatrix} = 0.$$

Suppose a symmetric matrix $\mathcal{P}_2$ is in the form of (31), we then obtain

$$\mathcal{A}\mathcal{P}_2 + \mathcal{P}_2\mathcal{A}^T + (I - \mathcal{J})\mathcal{B}\mathcal{B}^T(I - \mathcal{J}^T)$$
$$= \mathcal{A}(\mathcal{P}_1 + \beta\Pi) + (\mathcal{P}_1 + \beta\Pi)\mathcal{A}^T + (I - \mathcal{J})\mathcal{B}\mathcal{B}^T(I - \mathcal{J}^T)$$
$$= \mathcal{A}\mathcal{P}_1 + \mathcal{P}_1\mathcal{A}^T + (I - \mathcal{J})\mathcal{B}\mathcal{B}^T(I - \mathcal{J}^T) = 0,$$

which indicates that $\mathcal{P}_2$ is also a solution of (23).

That completes the proof. □

Based on the above lemma, we then can obtain the network controllability Gramian in the following theorem, despite that the solutions of the Lyapunov-like equation in (23) are not unique.

**Theorem 3.** *Consider the network system $\Sigma$ as in (2) and let $\mathcal{P}_a$ be an arbitrary solution of the Lyapunov-like equation in (22). Then, the network controllability Gramian $\mathcal{P}$ is given by*

$$\mathcal{P} = \mathcal{P}_a + \beta_a \Pi, \tag{38}$$

*where $\Pi$ is defined in (31) and*

$$\beta_a = -\sigma_D^{-2} \cdot \left( \begin{bmatrix} \mathbf{1}^T D, \mathbf{1}^T M \end{bmatrix} \mathcal{P}_a \begin{bmatrix} D\mathbf{1} \\ M\mathbf{1} \end{bmatrix} \right). \tag{39}$$

*Proof.* Consider $\mathcal{P}_a$ as an arbitrary solution of (22). Theorem 2 indicates that the controllability Gramian $\mathcal{P}$ is also a solution of (22). Then, by Lemma 2, there exists a scalar $\beta_a$ such that

$$\mathcal{P} = \mathcal{P}_a + \beta_a \Pi. \tag{40}$$

Next, we prove that $\beta_a$ satisfies (39) based on the definition of network controllability Gramian in (22). Denote a vector $\nu^T := \begin{bmatrix} \mathbf{1}^T D, \mathbf{1}^T M \end{bmatrix} \in \mathbb{R}^{1 \times 2n}$. Then, we have

$$\nu^T \mathcal{A} = \begin{bmatrix} \mathbf{1}^T D, \mathbf{1}^T M \end{bmatrix} \begin{bmatrix} \mathbf{0} & I \\ -M^{-1}L & -M^{-1}D \end{bmatrix} = 0, \tag{41}$$

where $\mathbf{1}^T L = 0$ is used. Since the power series expansion of $e^{\mathcal{A}t}$ is

$$e^{\mathcal{A}t} = I + \sum_{k=1}^{\infty} \frac{\mathcal{A}^k t^k}{k!}, \tag{42}$$

we have
$$\nu^T e^{\mathcal{A} t} = \nu^T. \tag{43}$$

For the matrix $\mathcal{J}$, from its definition (21), we have
$$\nu^T \mathcal{J} = \begin{bmatrix} \mathbf{1}^T D, \mathbf{1}^T M \end{bmatrix} \cdot \sigma_D^{-1} \begin{bmatrix} \mathbf{1}\mathbf{1}^T D & \mathbf{1}\mathbf{1}^T M \\ \mathbf{0}_{n\times n} & \mathbf{0}_{n\times n} \end{bmatrix} = \nu^T, \tag{44}$$

where $\sigma_D = \mathbf{1}^T D \mathbf{1}$ is used. In summary, from (43) and (44), we have
$$\nu^T(e^{\mathcal{A}t} - \mathcal{J}) = \nu^T I - \nu^T \mathcal{J} = 0. \tag{45}$$

This implies from the definition of network controllability Gramian in (22), $\nu^T \mathcal{P} = 0$, and consequenly
$$\nu^T \mathcal{P} \nu = 0. \tag{46}$$

By substituting (40) into (46), we obtain
$$\nu^T (\mathcal{P}_a + \beta_a \Pi)\nu = 0. \tag{47}$$

Observe that $\nu^T \Pi \nu = (\mathbf{1}^T D \mathbf{1})^2 = \sigma_D^2$, therefore,
$$\beta_a = -\sigma_D^{-2} \nu^T \mathcal{P}_a \nu, \tag{48}$$

which is equivalent to (39). □

**Remark 4.** *The proof of Theorem 3 implies that $\mathcal{P}$ satisfying (23) and (46) uniquely exists. Let both $\mathcal{P}_a$ and $\mathcal{P}_b$ satisfy (23) and (46). Condition (23) implies that for the network controllability Gramian $\mathcal{P}$, there exist scalar $\beta_a$ and $\beta_b$ such that*
$$\mathcal{P}_a + \beta_a \Pi = \mathcal{P} = \mathcal{P}_b + \beta_b \Pi. \tag{49}$$

*Moreover, (46) yields*
$$\beta_a \nu^{\mathrm{T}} \Pi \nu = 0 = \beta_b \nu^{\mathrm{T}} \Pi \nu. \tag{50}$$

*Since $\nu^{\mathrm{T}} \Pi \nu = \sigma_D^2$ is nonzero, we have $\beta_a = \beta_b = 0$. Therefore, we have $\mathcal{P}_a = \mathcal{P} = \mathcal{P}_b$.*

**Remark 5.** *The value of $\mathcal{P}$ does not depend on $\mathcal{P}_a$, which can be any symmetric matrix satisfying the Lyapunov-like equation. Besides, the network controllability Gramian $\mathcal{P}$ defined in (22) is positive semidefinite, since there exists a nonzero vector $\nu$ such that $\mathcal{P}\nu = 0$, and the dimension of the nonzero nullspace $\mathcal{P}$ is larger or equal to one.*

Theorem 3 provides an approach to obtain the second-order network controllability Gramian $\mathcal{P}$ without computing the integral in (22). First, we solve the Lyapunov-like equation in (23) and obtain an arbitrary solution. Then, applying (38) leads to the $\mathcal{P}$ matrix. In [29], [30], the algorithms to solve the Sylvester equation (29) are proposed for nonsingular $A$ and $B$ matrices. Based on them, it is not difficult to generalize the methods to the singular case, e.g., the Lyapunov-like equation in (23), just to acquire an arbitrary solution. In this paper, we will not further discuss the computation of the matrix equation (23) due to the limited space. In the following subsection, we adopt the second-order network controllability Gramian to design a efficient cluster selection algorithm.

*B. Hierarchical Clustering*

Which method is used for cluster selection generally determines the approximation quality of the reduced-order system. Therefore, it plays a crucial role in model reduction of a network system. Contrary to the existing algorithms in the literature, we propose a novel one that uses the $\mathcal{H}_2$-norms of transfer function discrepancy as the criterion to measure the *dissimilarities* of vertices and clusters those vertices with similar behaviors.

We first consider the transfer function of system $\Sigma$ in (2)

$$\eta(s) := (s^2 M + sD + L)^{-1} F \in \mathbb{R}(s)^{n \times m}, \tag{51}$$

and characterize the vertex behavior by the transfer function from external inputs to individual states, i.e., the behavior of the $i$-th vertex is captured by the transfer function

$$\eta_i(s) := \mathbf{e}_i^T \eta(s). \tag{52}$$

Then, the *dissimilarity* of vertices $i$ and $j$ is defined by

$$\mathcal{D}_{ij} = \|\eta_i(s) - \eta_j(s)\|_{\mathcal{H}_2}. \tag{53}$$

The boundedness of $\mathcal{D}_{ij}$ is implied by the following lemma.

**Lemma 3.** *Consider the second-order network system $\Sigma$ and add an external output: $y = H_s x + H_v \dot{x}$ with $H_s, H_v \in \mathbb{R}^{p \times n}$, then the input-output transfer function*

$$G(s) := (H_s + sH_v)(s^2 M + sD + L)^{-1} F \in \mathbb{R}(s)^{p \times m} \tag{54}$$

*is $\mathcal{H}_2$ norm-bounded if and only if $H_s \mathbf{1} = 0$ or $\mathbf{1}^T F = 0$.*

*Proof.* Denote $g(t)$ as the impulse response of $G(s)$, which is given by

$$g(t) = [H_s, H_v] e^{\mathcal{A}t} \mathcal{B} \in \mathbb{R}^{2p \times m} \tag{55}$$

with $\mathcal{A}$ and $\mathcal{B}$ defined in (5). Theorem 1 implies that $e^{\mathcal{A}t}$ is a bounded smooth function of $t$ and exponentially converges to a constant matrix $\mathcal{J}$ as $t \to \infty$. Therefore, the function $g(t)$ is *integrable* if and only if $\lim_{t \to \infty} g(t) = 0$.

From (18), the function $g(t)$ has an exponential convergence as follows

$$g(t) \to [H_s, H_v] \mathcal{J} \mathcal{B} = \sigma_D^{-1} H_s \mathbf{1} \mathbf{1}^T F. \tag{56}$$

Observe that $H_s \mathbf{1} \in \mathbb{R}^{p \times 1}$ and $\mathbf{1}^T F \in \mathbb{R}^{1 \times m}$. Therefore, we have $\lim_{t \to \infty} g(t) = 0$ if and only if $H_s \mathbf{1} = 0$ or $\mathbf{1}^T F = 0$. Notice that the function $g(t)$ being integrable equivalently implies $\mathbf{tr}\left[g(t)^T g(t)\right]$ is also integrable, and the $\mathcal{H}_2$-norm of $G(s)$ is presented as

$$\|G(s)\|_{\mathcal{H}_2} = \left(\int_0^\infty \mathbf{tr}\left[g(t)^T g(t)\right] dt\right)^{1/2}, \tag{57}$$

Thus, the $\mathcal{H}_2$-norm of $G(s)$ is bounded if and only if $\mathbf{tr}\left[g(t)^T g(t)\right]$ is integrable that is $g(t)$ is integrable, or equivalently, $H_s \mathbf{1} = 0$ or $\mathbf{1}^T F = 0$. □

Notice that $\eta_i(s) - \eta_j(s)$ is in form of $G(s)$ with $H_s = \mathbf{e}_i^T - \mathbf{e}_j^T$ and $H_v = 0$. Since $(\mathbf{e}_i^T - \mathbf{e}_j^T)\mathbf{1} = 0$ for all $i, j \in \{1, 2, \cdots, n\}$, the above lemma then implies that $\mathcal{D}_{ij}$ in (53) is always bounded for the network system $\Sigma$.

Now we define a *dissimilarity matrix* $\mathcal{D}$ whose entries are $\mathcal{D}_{ij}$. Clearly, $\mathcal{D}$ is nonnegative symmetric matrix with zero diagonal elements.

Computing matrix $\mathcal{D}$ poses a major challenge, especially for large-scale systems, while there are several methods, such as using the Riemann sum or linear matrix inequalities as in e.g., [31]. However, the proposed network controllability Gramian provides a more efficient computational method due to the following theorem.

**Theorem 4.** *Consider the input-output transfer function $G(s) := (H_s + sH_v)\eta(s)$ as in (54). If $H_s \mathbf{1} = 0$ or $\mathbf{1}^T F = 0$, then the $\mathcal{H}_2$-norm of $G(s)$ is computed by*

$$\|G(s)\|_{\mathcal{H}_2}^2 = \mathbf{tr}\left(H\mathcal{P}H^T\right). \tag{58}$$

where $H := [H_s, H_v]$ and $\mathcal{P}$ is the second-order network controllability Gramian defined in (22). Specifically, the relation between the dissimilarity measure $\mathcal{D}_{ij}$ and the network controllability Gramian $\mathcal{P}$ is given by

$$\mathcal{D}_{ij} = \sqrt{\mathbf{tr}\left( \begin{bmatrix} \mathbf{e}_{ij}^T, \mathbf{0}_{1\times n} \end{bmatrix} \mathcal{P} \begin{bmatrix} \mathbf{e}_{ij} \\ \mathbf{0}_{n\times 1} \end{bmatrix} \right)}. \tag{59}$$

*Proof.* From (57), the $\mathcal{H}_2$-norm of $G(s)$ is given by

$$\|G(s)\|_{\mathcal{H}_2}^2 = \mathbf{tr}\left( \int_0^\infty H e^{\mathcal{A}t} \mathcal{B}\mathcal{B}^T e^{\mathcal{A}^T t} H^T dt \right). \tag{60}$$

If $\mathbf{1}^T F = 0$, then

$$\mathcal{J}\mathcal{B} = \sigma_D^{-1} \begin{bmatrix} \mathbf{1}\mathbf{1}^T F \\ \mathbf{0}_{n\times m} \end{bmatrix} = \mathbf{0}_{2n\times m}, \tag{61}$$

and if $H_s \mathbf{1} = 0$,

$$H\mathcal{J} = \sigma_D^{-1} \begin{bmatrix} H_s \mathbf{1}\mathbf{1}^T D & H_s \mathbf{1}\mathbf{1}^T M \end{bmatrix} = \mathbf{0}_{p\times 2n}. \tag{62}$$

When either of (61) or (62) holds, we have

$$\begin{aligned} H\mathcal{P}H^T &= \int_0^\infty H(e^{\mathcal{A}t} - \mathcal{J})\mathcal{B}\mathcal{B}^T(e^{\mathcal{A}^T t} - \mathcal{J}^T)H^T dt \\ &= \int_0^\infty H e^{\mathcal{A}t} \mathcal{B}\mathcal{B}^T e^{\mathcal{A}^T t} H^T dt, \end{aligned} \tag{63}$$

Consequently, we obtain (58) by equation (60). Moreover, taking $H_s = \mathbf{e}_i^T - \mathbf{e}_j^T = \mathbf{e}_{ij}^T$ and $H_v = 0$ then yields (59), since $\mathbf{e}_{ij}^T \mathbf{1} = 0$. □

To compute the dissimilarity matrix $\mathcal{D}$, we fist calculate $\mathcal{P}$ by Theorem 3 and then just apply vector-matrix multiplication to obtain all the entries of $\mathcal{D}$.

The entries of $\mathcal{D}$ indicate the similarities of vertices. Based on the $\mathcal{D}$ matrix, we propose a hierarchical clustering algorithm to generate an appropriate network clustering for the system $\Sigma$. This approach links the pairs of vertices that are in close proximity and place them into binary clusters. Then, the newly formed clusters can be merged into larger clusters according to the *cluster dissimilarity*. The dissimilarity of clusters $\mathcal{C}_\mu$ and $\mathcal{C}_\nu$ is defined by

$$\delta(\mathcal{C}_\mu, \mathcal{C}_\nu) = \frac{1}{|\mathcal{C}_\mu| \cdot |\mathcal{C}_\nu|} \sum_{i\in \mathcal{C}_\mu} \sum_{j\in \mathcal{C}_\nu} \mathcal{D}_{ij}. \tag{64}$$

The notation $\delta(\mathcal{C}_\mu, \mathcal{C}_\nu)$ is characterized by the average dissimilarity between all pairs of vertices in the clusters $\mathcal{C}_\mu$ and $\mathcal{C}_\nu$.

The idea of hierarchical clustering has been extensively used in many fields, including pattern recognition, data compression, computer graphics, and process networks, see [16], [32], [33]. This paper is the first one that introduces this clustering algorithm to model reduction of network systems, and defines the distance by the norm of transfer functions. In hierarchical clustering, we first assign each vertex into a individual cluster and then merge two clusters into a single one if they have the least dissimilarity. Finally we can cluster the vertices into a binary, hierarchical tree, which is called *dendrogram*. The pseudocode of hierarchical clustering is described in Algorithm 1. Notice that Algorithm 1 is a greedy method.

---

**Algorithm 1** Hierarchical Clustering

---

**Input:** $M$, $D$, $L$ and $F$, model order $n$, desired order $r$

**Output:** $P$, $\hat{M}$, $\hat{D}$, $\hat{L}$

1: Compute the Gramian $\mathcal{P}$ by Theorem 3
2: Compute the dissimilarity matrix $\mathcal{D}$ by Theorem 4
3: Place each vertex into its own singleton cluster, that is $\mathcal{C}_i \leftarrow \{i\}$ for all $1 \leq i \leq n$
4: $k \leftarrow n$
5: **while** $k > r$ **do**
6:     Set $\delta_\mathrm{m}$ to be an arbitrary large number
7:     **for** $i = 1 : k-1$ and $j = 2 : i-1$ **do**
8:         Compute $\delta(\mathcal{C}_i, \mathcal{C}_j)$ by (64)
9:         **if** $\delta_\mathrm{m} > \delta(\mathcal{C}_i, \mathcal{C}_j)$ **then**
10:             $\mu \leftarrow i$, $\nu \leftarrow j$, $\delta_\mathrm{m} \leftarrow \delta(\mathcal{C}_i, \mathcal{C}_j)$
11:         **end if**
12:     **end for**
13:     Merge cluster $\mu$ and $\nu$ into a single cluster
14:     $k \leftarrow k - 1$
15: **end while**
16: Compute $P \in \mathbb{R}^{n \times r}$
17: $\hat{M} \leftarrow P^T M P$, $\hat{D} \leftarrow P^T D P$, $\hat{L} \leftarrow P^T L P$

---

**Remark 6.** *The network controllability Gramian analysis leads to a pair-wise distance notion of vertices, and the clustering algorithm is a simple consequence of it. We can also adapt other*

*clustering algorithms, such as iterative clustering, K-means clustering, or other greedy clustering strategies, to our problem. We choose hierarchical clustering because it can obtain a reduced network with small approximation error and low computational cost.*

*C. Error Analysis*

Now, we analyze the approximation error between the full-order and reduced-order system. First, we denote

$$\hat{\eta}(s) := P(s^2\hat{M} + s\hat{D} + \hat{L})^{-1}P^T F \in \mathbb{R}(s)^{n\times m} \tag{65}$$

as the transfer function of the reduced-order system (9) and $\hat{\eta}_i(s) := \mathbf{e}_i^T \hat{\eta}(s)$. Then the following lemma indicates the boundedness of the approximation error.

**Lemma 4.** *Consider the second-order network system $\Sigma$ in (2) and the reduced model $\hat{\Sigma}$ in (9) resulting from graph clustering. $\eta(s)$ and $\hat{\eta}(s)$ are the transfer functions defined in (51) and (65), respectively. Then, the following statements holds:*

1) $\|\eta_i(s) - \hat{\eta}_j(s)\|_{\mathcal{H}_2}$ *is bounded for any $i, j = \{1, 2, \cdots, n\}$.*
2) $\|\eta(s) - \hat{\eta}(s)\|_{\mathcal{H}_2}$ *is bounded.*

*Proof.* Note that the $\mathcal{H}_2$-norm of $\eta_i(s) - \hat{\eta}_j(s)$ is given by

$$\|\eta_i(s) - \hat{\eta}_j(s)\|^2_{\mathcal{H}_2} = \int_0^\infty \|\mathbf{e}_i^T \xi(t) - \mathbf{e}_j^T \hat{\xi}(t)\|_2^2 dt, \tag{66}$$

where $\xi(t)$ and $\hat{\xi}(t)$ are the impulse responses of $\Sigma$ and $\hat{\Sigma}$, respectively. Furthermore, both $\xi(t)$ and $\hat{\xi}(t)$ are bounded smooth functions of $t$, which exponentially converge to the same value. From Remark 2, we have $\lim_{t\to\infty} \xi(t) = \lim_{t\to\infty} \hat{\xi}(t) = \sigma_D^{-1}\mathbf{1}\mathbf{1}^T F$. Therefore,

$$\lim_{t\to\infty}\left[\mathbf{e}_i^T \xi(t) - \mathbf{e}_j^T \hat{\xi}(t)\right] = 0. \tag{67}$$

For bounded initial conditions $\xi_i(0)$ and $\hat{\xi}_j(0)$, the integral in (66) is bounded, i.e., $\|\eta_i(s) - \hat{\eta}_j(s)\|^2_{\mathcal{H}_2} < \infty$. It means that the norm of each row of $\eta(s) - \hat{\eta}(s)$ is finite, therefore, $\|\eta(s) - \hat{\eta}(s)\|_{\mathcal{H}_2}$ is also bounded. □

Now, we explore the method to compute the approximation error in terms of the $\mathcal{H}_2$-norm. For simplicity, we denote

$$\hat{\mathcal{A}} = \begin{bmatrix} \mathbf{0}_{r\times r} & I_r \\ -\hat{M}^{-1}\hat{L} & -\hat{M}^{-1}\hat{D} \end{bmatrix}, \quad \hat{\mathcal{B}} = \begin{bmatrix} \mathbf{0}_{r\times m} \\ \hat{M}^{-1}P^T F \end{bmatrix}. \tag{68}$$

for the reduced system $\hat{\Sigma}$ in (9). Since $P\mathbf{1}_r = \mathbf{1}_n$, $\mathbf{1}_r^T \hat{D} \mathbf{1}_r = \mathbf{1}_r^T P^T DP\mathbf{1}_r = \mathbf{1}_n^T D\mathbf{1}_n$, the convergence matrix of $\hat{\Sigma}$ is given by

$$\hat{\mathcal{J}} = \sigma_D^{-1} \cdot \begin{bmatrix} \mathbf{1}_r \mathbf{1}_r^T \hat{D} & \mathbf{1}_r \mathbf{1}_r^T \hat{M} \\ \mathbf{0}_{r \times r} & \mathbf{0}_{r \times r} \end{bmatrix} \tag{69}$$

with $\sigma_D = \mathbf{1}_n^T D \mathbf{1}_n$.

Next, we consider the following error system:

$$\mathbf{\Sigma_e} : \begin{cases} \dot{\omega} = \mathcal{A}_e \omega + \mathcal{B}_e u, \\ \delta = \mathcal{C}_e \omega, \end{cases} \tag{70}$$

where

$$\mathcal{A}_e = \begin{bmatrix} \mathcal{A} & 0 \\ 0 & \hat{\mathcal{A}} \end{bmatrix}, \mathcal{B}_e = \begin{bmatrix} \mathcal{B} \\ \hat{\mathcal{B}} \end{bmatrix}, \mathcal{C}_e = \begin{bmatrix} I_n & \mathbf{0}_{n \times n} - P & \mathbf{0}_{n \times r} \end{bmatrix}.$$

Then, the approximation error between the full-order and reduced-order system is equivalent to computing $\|\mathbf{\Sigma_e}\|_{\mathcal{H}_2}$.

Lemma 4 guarantees that, by clustering-based projection, the approximation error between the full-order and reduced-order models is bounded. Now, we exploit the method to compute the errors $\|\Sigma - \hat{\Sigma}\|_{\mathcal{H}_2}$.

To this end, we define the *coupling network controllability Gramian* of system $\mathbf{\Sigma_e}$, which is formulated as

$$\mathcal{P}_\mathsf{x} = \int_0^\infty (e^{\mathcal{A}t} - \mathcal{J})\mathcal{B}\hat{\mathcal{B}}^T (e^{\hat{\mathcal{A}}^T t} - \hat{\mathcal{J}}^T) dt \in \mathbb{R}^{2n \times 2r}. \tag{71}$$

The following lemma provides a method to obtain $\mathcal{P}_\mathsf{x}$ without integration.

**Lemma 5.** *Consider the error system $\mathbf{\Sigma_e}$ in (70) and its coupling network controllability Gramian is computed by*

$$\mathcal{P}_\mathsf{x} = \tilde{\mathcal{P}}_\mathsf{x} + \beta_\mathsf{x} \Pi_\mathsf{x}, \tag{72}$$

*where $\tilde{\mathcal{P}}_\mathsf{x}$ is an arbitrary solution of the following Sylvester-like equation*

$$\mathcal{A}\tilde{\mathcal{P}}_\mathsf{x} + \tilde{\mathcal{P}}_\mathsf{x} \hat{\mathcal{A}} + \mathcal{B}\hat{\mathcal{B}}^T = 0, \tag{73}$$

*and $\beta_\mathsf{x}$ is a scalar constant given by*

$$\beta_\mathsf{x} = -\sigma_D^{-2} \cdot \begin{bmatrix} \mathbf{1}_n^T D, \mathbf{1}_n^T M \end{bmatrix} \tilde{\mathcal{P}}_\mathsf{x} \begin{bmatrix} \hat{D}\mathbf{1}_r \\ \hat{M}\mathbf{1}_r \end{bmatrix} \tag{74}$$

*with $\sigma_D = \mathbf{1}_n^T D \mathbf{1}_n$.*

*Proof.* By similar reasoning as in the proofs of Theorem 2, Lemma 2 and Theorem 3, the following results can be obtained:

First, $\mathcal{P}_\mathsf{x}$ is the solution of the matrix equation in (73).

Second, both $\tilde{\mathcal{P}}_\mathsf{x}$ and $\mathcal{P}_\mathsf{x}$ are the solutions of equation (73) if only if they satisfy

$$\mathcal{J}(\tilde{\mathcal{P}}_\mathsf{x} - \mathcal{P}_\mathsf{x})\hat{\mathcal{J}}^T = \tilde{\mathcal{P}}_\mathsf{x} - \mathcal{P}_\mathsf{x}, \tag{75}$$

or equivalently, $\mathcal{P}_\mathsf{x}$ can be expressed as

$$\mathcal{P}_\mathsf{x} = \tilde{\mathcal{P}}_\mathsf{x} + \beta_\mathsf{x} \Pi_\mathsf{x}, \tag{76}$$

with $\Pi_\mathsf{x} := \begin{bmatrix} \mathbf{1}_n \mathbf{1}_r^T & \mathbf{0}_{n \times r} \\ \mathbf{0}_{n \times r} & \mathbf{0}_{n \times r} \end{bmatrix}$ and $\beta_\mathsf{x}$ a scalar constant.

Third, $\mathcal{P}_\mathsf{x}$ satisfies

$$\begin{bmatrix} \mathbf{1}_n^T D, \mathbf{1}_n^T M \end{bmatrix} \mathcal{P}_\mathsf{x} \begin{bmatrix} \hat{D} \mathbf{1}_r \\ \hat{M} \mathbf{1}_r \end{bmatrix} = 0. \tag{77}$$

Note that

$$\begin{bmatrix} \mathbf{1}_n^T D, \mathbf{1}_n^T M \end{bmatrix} \Pi_\mathsf{x} \begin{bmatrix} \hat{D} \mathbf{1}_r \\ \hat{M} \mathbf{1}_r \end{bmatrix} = \mathbf{1}_n^T D \mathbf{1}_n \mathbf{1}_r^T \hat{D} \mathbf{1}_r = \sigma_D^2. \tag{78}$$

Therefore, from (76) and (77), we obtain the expression of $\beta_\mathsf{x}$ as in (74). $\square$

Based on the coupling network controllability Gramian, the approximation error between the full-order and reduced-order system, i.e., $\|\mathbf{\Sigma_e}\|_{\mathcal{H}_2}$ is obtained as follows.

**Theorem 5.** *Consider the second-order network system $\mathbf{\Sigma}$ in (2) and the reduced model $\hat{\mathbf{\Sigma}}$ in (9) resulting from graph clustering. Then the error between $\mathbf{\Sigma}$ and $\hat{\mathbf{\Sigma}}$ in terms of the $\mathcal{H}_2$-norm is computed by*

$$\|\mathbf{\Sigma} - \hat{\mathbf{\Sigma}}\|_{\mathcal{H}_2} = \sqrt{\mathbf{tr}\left(\mathcal{C}_\mathbf{e} \begin{bmatrix} \mathcal{P}_n & \mathcal{P}_\mathsf{x} \\ \mathcal{P}_\mathsf{x}^T & \mathcal{P}_r \end{bmatrix} \mathcal{C}_\mathbf{e}^T\right)}, \tag{79}$$

*where $\mathcal{C}_\mathbf{e}$ is defined in (70), and $\mathcal{P}_n \in \mathbb{R}^{2n \times 2n}$ and $\mathcal{P}_r \in \mathbb{R}^{2r \times 2r}$ are the second-order network controllability Gramians of the full-order system $\mathbf{\Sigma}$ and reduced-order system $\hat{\mathbf{\Sigma}}$, respectively. $\mathcal{P}_\mathsf{x}$ is the coupling network controllability Gramian of the error system (70).*

*Proof.* We extend the concept of network controllability Gramian to the error system $\mathbf{\Sigma_e}$:

First, the convergence matrix of system $\Sigma_{\mathbf{e}}$ is given by

$$\mathcal{J}_{\mathbf{e}} = \lim_{t\to\infty} e^{\mathcal{A}_{\mathbf{e}} t} = \begin{bmatrix} \mathcal{J} & 0 \\ 0 & \hat{\mathcal{J}} \end{bmatrix}. \qquad (80)$$

Second, the network controllability Gramian of system $\Sigma_{\mathbf{e}}$ is defined by

$$\mathcal{P}_{\mathbf{e}} = \int_0^\infty (e^{\mathcal{A}_{\mathbf{e}} t} - \mathcal{J}_{\mathbf{e}}) \mathcal{B}_{\mathbf{e}} \mathcal{B}_{\mathbf{e}}^T (e^{\mathcal{A}_{\mathbf{e}}^T t} - \mathcal{J}_{\mathbf{e}}^T) dt, \qquad (81)$$

which can partitioned as

$$\mathcal{P}_{\mathbf{e}} = \begin{bmatrix} \mathcal{P}_n & \mathcal{P}_{\times} \\ \mathcal{P}_{\times}^T & \mathcal{P}_r \end{bmatrix}. \qquad (82)$$

Note that $\mathcal{C}_{\mathbf{e}} \mathcal{J}_{\mathbf{e}} \mathcal{B}_{\mathbf{e}} = 0$, since

$$\mathcal{C}_{\mathbf{e}} \mathcal{J}_{\mathbf{e}} \mathcal{B}_{\mathbf{e}} = \begin{bmatrix} [I_n, \mathbf{0}_n] \mathcal{J} & [-P, \mathbf{0}_r] \hat{\mathcal{J}} \end{bmatrix} \mathcal{B}_{\mathbf{e}}$$

$$= \sigma_D^{-1} \begin{bmatrix} \mathbf{1}_n \mathbf{1}_n^T D & \mathbf{1}_n \mathbf{1}_n^T M & -P \mathbf{1}_r \mathbf{1}_r^T \hat{D} & -P \mathbf{1}_r \mathbf{1}_r^T \hat{M} \\ \mathbf{0}_{n\times n} & \mathbf{0}_{n\times n} & \mathbf{0}_{n\times r} & \mathbf{0}_{n\times r} \end{bmatrix} \cdot \begin{bmatrix} \mathbf{0}_{n\times m} \\ M^{-1} F \\ \mathbf{0}_{r\times m} \\ \hat{M}^{-1} P^T F \end{bmatrix}$$

$$= \begin{bmatrix} \mathbf{1}_n \mathbf{1}_n^T F - P \mathbf{1}_r \mathbf{1}_r^T P^T F \\ \mathbf{0}_{n\times m} \end{bmatrix} = \mathbf{0}_{2n\times m}.$$

Therefore,

$$\begin{aligned} \mathcal{C}_{\mathbf{e}} \mathcal{P}_{\mathbf{e}} \mathcal{C}_{\mathbf{e}}^T &= \int_0^\infty \mathcal{C}_{\mathbf{e}} (e^{\mathcal{A}_{\mathbf{e}} t} - \mathcal{J}_{\mathbf{e}}) \mathcal{B}_{\mathbf{e}} \mathcal{B}_{\mathbf{e}}^T (e^{\mathcal{A}_{\mathbf{e}}^T t} - \mathcal{J}_{\mathbf{e}}^T) \mathcal{C}_{\mathbf{e}}^T dt \\ &= \int_0^\infty \mathcal{C}_{\mathbf{e}} e^{\mathcal{A}_{\mathbf{e}} t} \mathcal{B}_{\mathbf{e}} \mathcal{B}_{\mathbf{e}}^T e^{\mathcal{A}_{\mathbf{e}}^T t} \mathcal{C}_{\mathbf{e}}^T dt. \end{aligned} \qquad (83)$$

Finally, we have $\|\Sigma - \hat{\Sigma}\|_{\mathcal{H}_2} = \|\Sigma_{\mathbf{e}}\|_{\mathcal{H}_2} = \sqrt{\mathbf{tr}\,(\mathcal{C}_{\mathbf{e}} \mathcal{P}_{\mathbf{e}} \mathcal{C}_{\mathbf{e}}^T)}.$ □

Now, an extension of the clustering-based method to *velocity-based model reduction* for second-order network system is discussed, where the closeness of $\dot{x}_i$ and $\dot{x}_j$ with different $i, j$ is considered. For a given system $\Sigma$ as in (2), we intend to find a system

$$\hat{\Sigma}_v : \begin{cases} \hat{M} \ddot{z} + \hat{D} \dot{z} + \hat{L} z = P^T F u, \\ v = P \dot{z}, \end{cases} \qquad (84)$$

such that the state $\dot{x}(t)$ in (2) is approximated by $v(t)$ in $\hat{\Sigma}_v$. To this end, the network controllability Gramian $\mathcal{P}$ in (22) is used. Denote

$$\zeta(s) := s(s^2 M + sD + L)^{-1} F \quad \text{and} \quad \zeta_i(s) := \mathbf{e}_i^T \zeta(s). \qquad (85)$$

The behavior of the $i$-th vertex is captured by the transfer function from external inputs to the velocity of the $i$-th vertex $\dot{x}(t)$. Then, we consider a velocity dissimilarity matrix $\mathcal{D}^v$ with the $(i,j)$-th entry as

$$\mathcal{D}^v_{ij} = \|\zeta_i(s) - \zeta_j(s)\|_{\mathcal{H}_2} \tag{86}$$

Here, the dissimilarity between two vertices is measured by the velocity difference over time. Similar to (59), $\mathcal{D}^v_{ij}$ is computed by

$$\|\mathcal{D}^v_{ij}\|_{\mathcal{H}_2} = \sqrt{\mathbf{tr}\left(\begin{bmatrix}\mathbf{0}_{1\times n}, \mathbf{e}^T_{ij}\end{bmatrix} \mathcal{P} \begin{bmatrix}\mathbf{0}_{n\times 1} \\ \mathbf{e}_{ij}\end{bmatrix}\right)}. \tag{87}$$

Using the matrix $\mathcal{D}^v$, the hierarchical clustering algorithm is also applicable for *velocity-based model reduction*. To estimate the approximation error, we just replace $\mathcal{C}_e$ in Theorem 5 by

$$\mathcal{C}^v_e = \begin{bmatrix}\mathbf{0}_{n\times n} & I_n & \mathbf{0}_{n\times r} & -P\end{bmatrix}. \tag{88}$$

## V. SMALL-WORLD NETWORK EXAMPLE

In this section, we demonstrate the feasibility of our model reduction method by a simulation. We generate a mass-damper-spring system evolving on a undirected connected network with 70 vertices, see Fig. 2. The blue, red and yellow segments present the edges connecting by springs, dampers and both of them, respectively.

The masses are set by

$$M_{ii} = (i \bmod 10) + 1, \tag{89}$$

where **mod** presents a modulo operation. In Fig. 2, the bigger size of a vertex means it has a larger mass. The topologies of spring and damper couplings are generated by Watts-Strogatz model [34], which is a random graph generator producing graphs with small-world properties. Furthermore, the damper on each vertex is set to be proportional to its mass. we add 5 inputs to the network, and the input matrix $F$ is randomized as a 70-by-5 matrix, whose entries are in the range of $[-1, 1]$.

We apply the hierarchical clustering algorithm to reduce the full-order second-order network system. The two clusters with the nearest distance are merged into a single one and finally, Algorithm 1 will group the vertices into a binary, hierarchical tree, called *dendrogram*, see Fig. 3. The dendrogram is fairly straightforward to interpret the result of graph clustering: The bottom vertical lines are called leaves, which represent the vertices on graph. Besides, each fusion of two clusters is indicated by the splitting of a vertical line into two branches, and the horizontal position of the split, shown by the short horizontal bar, reads the similarities between the two clusters. In Fig. 3, we use five different colors to show the result of graph clustering with five clusters.

The resulting reduced network systems are shown in Fig. 4, where simplified networks with different number of vertices are presented. We find that the simplified network system with lower dimension trends to have more edges simultaneously connected by springs and dampers.

Next, we compare our hierarchical clustering algorithm with other clustering strategies to illustrate that hierarchical clustering is effective in obtaining a reduced-order network system with smaller errors. The two additional strategies we use for the comparison are described as follows.

- Random clustering randomly assigns $n$ vertices in set $\mathcal{V}$ into $r$ nonempty subsets.
- Simple greedy clustering aggregates the vertices if they have smaller pair-wise dissimilarities. More specifically, it first clusters the most similar two vertices and then the pairs of vertices with the second smallest dissimilarity. In the following steps, it recursively aggregates the vertices with bigger and bigger dissimilarity until $r$ clusters are obtained. At each step, two clusters are unified if they have intersections.

Fig. 5 depicts the comparison of three strategies in their approximation errors. The random clustering is performed for 50 times, and the average of the approximation errors is plotted in Fig. 5. Generally, the errors obtained by the different clustering strategies decrease as the reduced order $r$ increases. However, it is clear that the hierarchical clustering algorithm has better performance than the other two strategies. When $r = 5$, the approximation error obtained by hierarchical clustering is $\|\Sigma - \hat{\Sigma}\|_{\mathcal{H}_2} = 0.5967$, which implies that behaviors of the full-order model can be well-approximated.

We implement this numerical experiment by Matlab 2016a in the environment of 64-bit operating system with Intel Core i5-3470 CPU @ 3.20GHz, RAM 8.00 GB. To find the fifth dimensional simplified model, it costs 1.1656s, while the time of computing the Gramian is 1.1167s. Therefore, the time consumption is mainly taken by the first step of Algorithm 1, and once the network controllability Gramian is obtained, the hierarchical clustering can be processed rapidly.

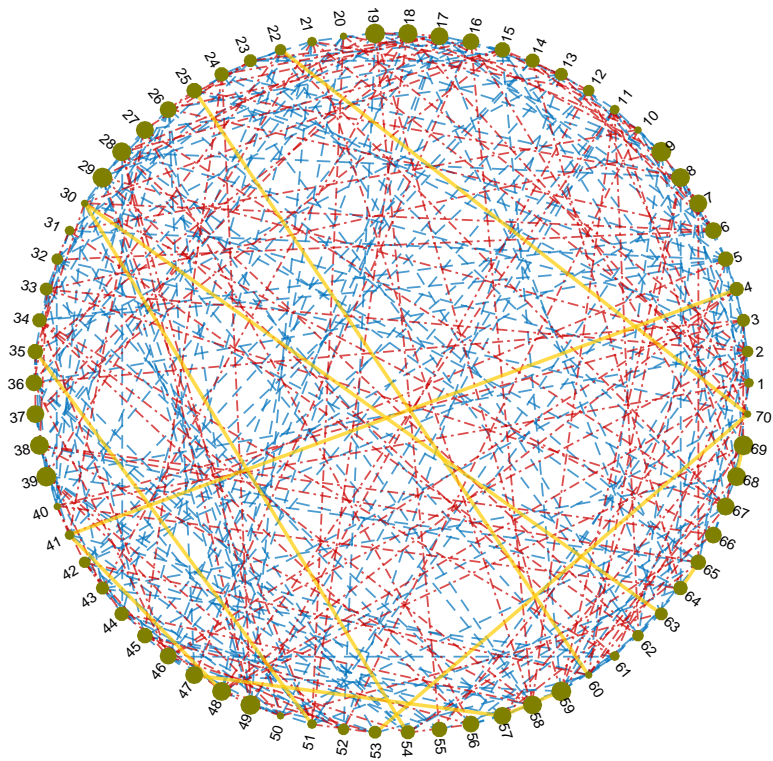

Fig. 2: Original Second-order networks with 70 vertices

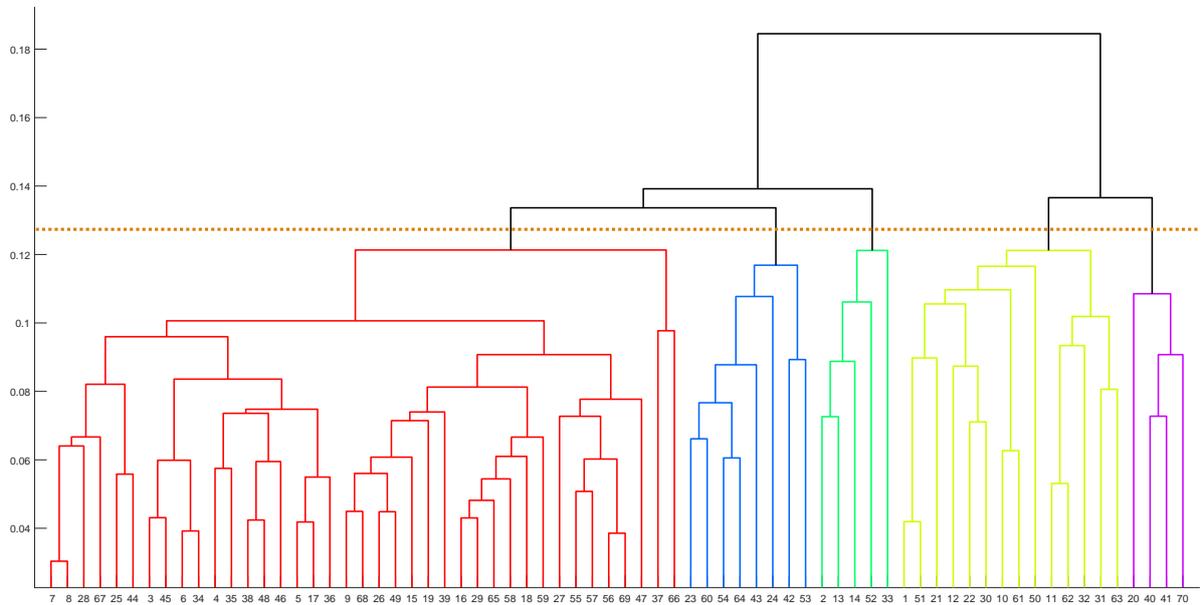

Fig. 3: Dendrogram depicting the graph clustering, where the horizontal axis are labeled by vertex numberings, while the vertical axis represents the dissimilarity between clusters.

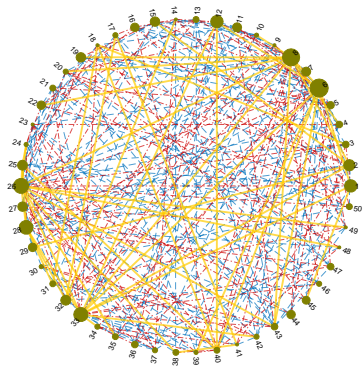

(a) Network with 50 clusters

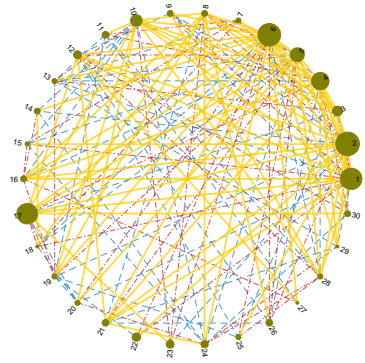

(b) Network with 30 clusters

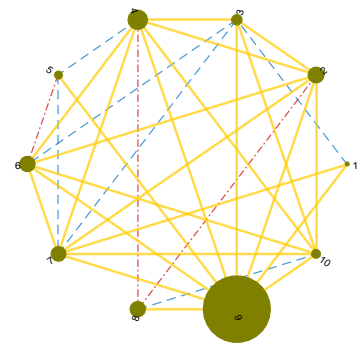

(c) Network with 10 clusters

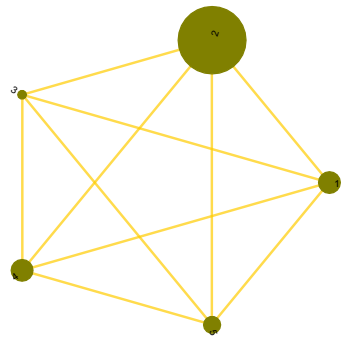

(d) Network with 5 clusters

Fig. 4: Reduced network with different numbers of clusters

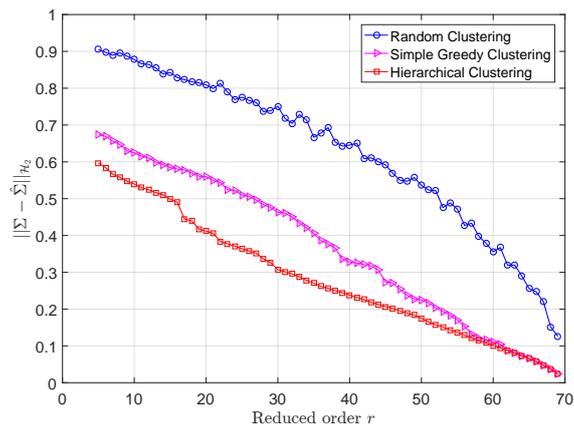

Fig. 5: Approximation error comparisons of hierarchical clustering algorithm with random and simple greedy clustering strategies, where the curve of random clustering is plotted based on the mean of 50 times experiments.

## VI. Conclusion

Based on graph clustering, we have developed a model reduction method for interconnected second-order systems. A hierarchical clustering algorithm is proposed to find an appropriate clustering such that the vertices with similar input responses are merged. Then, a projection using cluster matrix is applied to yield lower-dimensional network model. It is verified that such reduced system preserves network structures. Besides, we introduce the network controllability Gramian for the computation of $\mathcal{H}_2$-norms, which improve the feasibility of our algorithm. Finally, the efficiency of the proposed method has been illustrated by an experiment.

It is worth mentioning that although we consider a linear second-order system as in (2), the proposed method can be extended to different types of consensus networks. Our future work includes extensions to nonlinear networks and to network systems with subsystems of higher-order linear dynamics.